\documentclass[prl,superscriptaddress,floatfix,amsmath,footinbib,amssymb,twocolumn]{revtex4}
\usepackage{amssymb}
\usepackage{amsmath}
\usepackage{amsfonts}
\usepackage{bbold}
\usepackage{tikz}
\usepackage{mathtools}
\usepackage{soul}
\usepackage{bm}
\usepackage{ulem}

\usepackage{epsfig}
\usepackage{t1enc}
\usepackage{soul}
\usepackage{color}

\usepackage{pgfplots}
\pgfplotsset{compat=1.15}

\usepackage{mathrsfs}
\usepackage{url}
\usepackage[all]{xy}
\usetikzlibrary{calc,patterns,angles,quotes}
\usetikzlibrary{arrows.meta}

\newcommand{\superspace}{twin space}
\newcommand{\superstate}{twin state}

\newsavebox\mypic
\savebox\mypic{\, \begin{tikzpicture} \draw[->, bend left=20] (0,0) to node[above] {\scriptsize{$\Lambda$}} (0.5,0);  \end{tikzpicture} \,}

\DeclareMathOperator{\sgn}{sgn}
\DeclareMathOperator{\Tr}{Tr}
\definecolor{Jerzy}{rgb}{1.0, 0.65, 0.79}
\definecolor{Krzys}{rgb}{0.1, 0.6, 0.4}
\definecolor{Szymon}{rgb}{1.0, 0.7, 0.0}
\usepackage[final,stretch=30,shrink=40]{microtype}

\begin{document}

\title{Covariant quantum field theory of tachyons}

\author{Jerzy Paczos}
\affiliation{Department of Physics, Stockholm University, SE-106 91 Stockholm, Sweden}

\author{Kacper Dębski}
\affiliation{Institute of Theoretical Physics, University of Warsaw, Pasteura 5, 02-093 Warsaw, Poland}

\author{Szymon Cedrowski}
\affiliation{Institute of Theoretical Physics, University of Warsaw, Pasteura 5, 02-093 Warsaw, Poland}

\author{Szymon Charzyński}
\affiliation{Department of Mathematical Methods in Physics, University of Warsaw, Pasteura 5, 02-093 Warsaw, Poland}

\author{Krzysztof Turzyński}
\affiliation{Institute of Theoretical Physics, University of Warsaw, Pasteura 5, 02-093 Warsaw, Poland}

\author{Artur Ekert}
\affiliation{Centre for Quantum Technologies, National University of Singapore, 3 Science Drive 2, 117543 Singapore, Singapore}
\affiliation{Mathematical Institute, University of Oxford, Woodstock Road, Oxford OX2 6GG, United Kingdom}
\affiliation{Okinawa Institute of Science and Technology, Onna, Okinawa 904-0495, Japan}

\author{Andrzej Dragan}
\email{dragan@fuw.edu.pl}
\affiliation{Institute of Theoretical Physics, University of Warsaw, Pasteura 5, 02-093 Warsaw, Poland}
\affiliation{Centre for Quantum Technologies, National University of Singapore, 3 Science Drive 2, 117543 Singapore, Singapore}

\date{\today}

\begin{abstract}
Three major misconceptions concerning quantized tachyon fields: the energy spectrum unbounded from below, the frame-dependent and unstable vacuum state, and the non-covariant commutation rules, are shown to be a result of misrepresenting the Lorentz group in a too small Hilbert space. By doubling this space we establish an explicitly covariant framework that allows for the proper quantization of the tachyon fields eliminating all of these issues. Our scheme that is derived to maintain the relativistic covariance also singles out the two-state formalism developed by Aharonov {\it et al.} \cite{Aharonov1964} as a preferred interpretation of the quantum theory.
\end{abstract}

\maketitle



\noindent {\it ---Introduction.} Tolman's hypothesis \cite{Tolman1917} that superluminal particles lead to causality paradoxes \cite{Pirani1970,Parmentola1971,Landsberg1972,Goldhaber1975,Barrowes1977,Basano1977,Maund1979,Basano1980,Antippa1972}, has recently been challenged by Dragan and Ekert \cite{Dragan2020} who showed that the tachyons do not create logical paradoxes, but only disturbances of causality akin to those known in quantum theory. An important mathematical difficulty, which prevented the extension of special relativity to include superluminal phenomena in spacetimes of dimensions greater than 1+1 \cite{Machildon1983}, was also recently overcome \cite{Dragan2023}, catalyzing the discussion around the claim that tachyons, being the {\it enfant terrible} of physics, are not definitively ruled out from its domain \cite{Gavassino_2022, Grudka_2022, DelSanto_2022, grudka2023superluminal, Dragan_2022reply, dragan2022reply2, horodecki2023quantum}.

Since the 1960s, physicists have attempted to propose a quantum theory of tachyons \cite{Feinberg1967, Arons1968, Dhar1968, Susskind} and experimentally detect them \cite{Alvager1968,Davis1969, CLAY1974,Murthy1971, Smith1977, Baltay1970}. Chodos and Kostelecky \cite{Chodos} even suggested that at least one known neutrino species might be a tachyon, prompting experimental tests and subsequent studies \cite{Jentschura2012,Diaz2014,Cohen2011, Antonello2012, PhysRevLett.122.091803,Stecker2014,Jentschura2016,EHRLICH20131,Ehrlich2019,Schwartz2018,Ehrlich_sym14061198}, in particular the infamous claim of a faster-than-light neutrino in the OPERA experiment \cite{Adam2012}. Within the framework of string theory, the existence of tachyons is also well-recognized, though they are typically regarded as unwelcome artifacts \cite{Green,Banks,Lifschytz,Sen_2002TachyonMatter,Sen2002fieldtheoryoftachyonmatter,Sen1999}. Tachyons have also been studied in the context of the Casimir effect \cite{Nesterenko1991,Ostrowski2005}, statistical mechanics and thermodynamics  \cite{Bhattacharya2012,Dawe1989,Chattopadhyay, Sulehri2021, Kowalski2007,TOMASCHITZ2007558,Mrowczynski1984}. Studies on the tachyon field's potential applications in cosmology reveal diverse possibilities \cite{Alexander2001,Das,Sen,Gibbons_2003,Sen_2002,MAZUMDAR2001101,SARANGI2002185}, notably in mechanisms like the "rolling of the tachyon" 
\cite{admanabhan2002,FROLOV2002,LevKofman_2002,FAIRBAIRN20021,SHIU20026,GaryFelder_2002,JamesCline_2002,KIM2003111} and "transient tachyonic instabilities" enhancing cosmological perturbation amplitudes. Tachyons are often associated with instabilities, as highlighted in several studies \cite{Felder2001,Frusciante_2019,Acatrinei}, in particular, in ``tachyonic preheating'', the phenomenon of rapid energy transfer due to tachyonic instability \cite{Felder2001_2,Copeland,Koivunen_2022,Tomberg_2021,He_2021,AlejandroArrizabalaga_2004,Enqvist_2005,Dufaux_2009,Karam_2021}.  It has also been pointed out \cite{Dragan2023} that fields with ``negative mass square'', i.e., tachyon fields, are integral to models of spontaneous symmetry breaking, such as the Higgs mechanism \cite{Srednicki}.

Unfortunately, all attempts for further discussion about tachyons are held by serious mathematical difficulties that prevent their covariant description within the framework of quantum field theory (QFT). Since the earliest attempts \cite{Feinberg1967, Arons1968, Dhar1968, Susskind}, at least three such problems remain unsolved: the energy spectrum is unbounded from below, Lorentz boosts alter an already unstable vacuum and the commutation rules are not covariant with respect to Lorentz boosts.

In this work, we show that these issues stem from the improper representation of the Lorentz group in a too-small Hilbert space. After a natural extension of this space, problems with the quantization of tachyon fields disappear and it is possible to construct a theory with a stable and relativistically invariant tachyon vacuum and a lower-bounded energy spectrum. Notably, our findings align with the two-state formalism in quantum mechanics, introduced by Aharonov, Bergmann, and Lebowitz \cite{Aharonov1964} to guarantee time-reversibility of the measurement process. While this formalism was only an exotic interpretation of non-relativistic quantum mechanics, in quantum field theory of tachyons, a similar approach turns out to be a necessary element guaranteeing the relativistic invariance \footnote{It must be pointed out that our work, while it deals with quantum field theory for tachyons, does not address the complex interpretational questions linked to the two-state vector formalism. Issues such as the interpretation of weak measurements or the validity of the product rule remain outside the scope of our study. In quantum field theory, defining certain observables, like the number of particles in a finite space, can be problematic \cite{Kialka2018}, but our use of the two-state formalism is confined to its utility in computing probability amplitudes for processes using $S$-matrix elements, and does not extend to these broader interpretative concerns.}.

Previous attempts at tachyon quantization have proven unsuccessful. Tanaka's early study \cite{Tanaka1960} on virtual tachyons lacked a complete covariant description. Feinberg's approach \cite{Feinberg1967} aiming to preserve relativistic covariance by swapping the scalar theory's commutation relations with anti-commutation relations also failed. Arons and Sudarshan \cite{Arons1968}, through an explicit derivation of the Bogolyubov transformation of the vacuum state, demonstrated that Feinberg's attempt did not succeed and that Lorentz invariance was still compromised. Arons, Sudarshan, and Dhar \cite{Arons1968,Dhar1968} proposed their own remedy, maintaining the Lorentz-invariance but adding a frequency index to the annihilation operators, which caused some annihilation operators to function as creation operators. However, they also required the vacuum state to be 'annihilated' by these effective creation operators, which is mathematically impossible. As a result, their entire construction hinged on self-contradictory assumptions. In another attempt, Schwartz \cite{Schwartz} suggested a novel quantization method for the tachyonic field distinguishing two space-like directions and treating a third one as a function of frequency. He also made another observation \cite{Schwartz2018} analogous to Sudarshan's reinterpretation of the emission/absorption process \cite{Bilaniuk1962} that the space of single-particle states is not Lorentz-invariant and should be enlarged. We also consider this observation as our starting point, but we approach it more rigorously, making it the foundation of our relativistically invariant framework of quantum field theory of tachyons and their interactions with other fields.


\noindent {\it ---Lorentz covariance of quantum tachyonic fields.} The action of the Lorentz group for standard quantum fields is represented in the Fock space $\cal F$ constructed as a direct sum of definite particle number subspaces:
\begin{align}
\label{fockspace}
{\cal F} \equiv  \bigoplus_{n=0}^\infty S\left({\cal H}^{\otimes n}\right),	
\end{align}
where ${\cal H}$ is a one-particle Hilbert space and $S$ is the symmetrization operator. A representation of the Lorentz group can be characterized by defining the action of the group elements on the one-particle subspace ${\cal H}$ only because Lorentz boosts do not change the number of particles, thereby keeping ${\cal H}$ invariant.

\begin{figure}[h]
  \centering
  \begin{tikzpicture}[scale=1]
    \draw[->] (-1,-1) -- (1.5,-1) node[below right] {$\smash{x}$};
    \draw[->] (-1,-1) -- (-1,1.5) node[left] {$ct$};
    \node at (1.2,1.4) {a)};

	\draw (-0.5,-0.7) -- (0,0);
	\draw (0,0) -- (0.1,1.1);
	\draw[dashed] (0,0) -- (1.2,0.3);

    \draw[->, bend left=20] (2.2,0.3) to node[above] {$\Lambda$} (3.3,0.3);

    \begin{scope}[xshift=5.2cm]
	\draw[->] (-1,-1) -- (1.5,-1) node[below right] {$\smash{x'}$};
    \draw[->] (-1,-1) -- (-1,1.5) node[left] {$ct'$};
    \node at (1.2,1.4) {b)};

    \draw (-0.31,-0.7) -- (0,0);
    \draw (0,0) -- (-0.35,1.1);
    \draw[dashed] (0,0) -- (1.17,-0.2);

    \end{scope}
  \end{tikzpicture}
  \caption{\label{absorptionemission} 
  An emission of a positive-energy tachyon (a) in one reference frame is Lorentz-transformed by a boost $\Lambda$ into an emission of a negative-energy tachyon (b) backward in time, which can be reinterpreted as a regular absorption of a positive-energy antitachyon forward in time, in the boosted reference frame. The worldlines of tachyons are shown as dashed lines.} 
\end{figure}
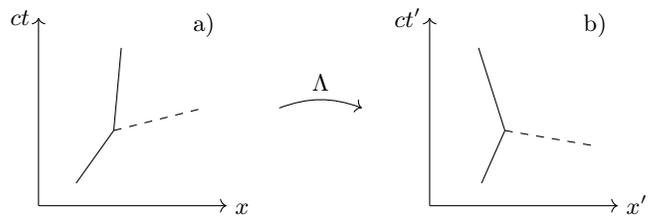

Contrarily, the case of tachyons differs. Classically, a tachyon is a particle moving along a space-like trajectory and characterized by a space-like energy-momentum four-vector tangent to that trajectory \cite{Dragan2020a,Dragan2023}. Therefore, a Lorentz boost can transform a positive-energy tachyon moving forward in time into a negative-energy tachyon moving backward in time. This elementary fact has far-reaching consequences for the quantum theory of tachyons. To illustrate this point, consider a hypothetical process of a subliminal particle emitting a positive-energy tachyon as depicted in Fig.~\ref{absorptionemission}a). There exists another inertial observer, for whom the same process involves the emission of a negative-energy tachyon moving backward in time, as shown in Fig.~\ref{absorptionemission}b). In this reference frame, the process can be reinterpreted as an absorption of a positive-energy anti-tachyon moving forward in time \cite{Arons1968}. 
As a result, both the number of tachyons in the input state and the number of tachyons in the output state turn out to be frame-dependent.

This scenario indicates that a Lorentz boost can \textcolor{Jerzy} interchange an output state of the tachyon field with an input state. Therefore the Lorentz group's action for tachyonic fields should not be represented in the space of input states ${\cal F}$, which lacks boost invariance. Instead, the Lorentz group should be represented in the \superspace\ ${\cal F}\otimes{\cal F}^\star$, containing both input and output states (with ${\cal F}^\star$ denoting the space dual to ${\cal F}$). This \superspace\ characterizes both the past and the future of the field and is the smallest Hilbert space preserved under the  Lorentz (as well as Poincare) group's action for tachyons. Previous issues with the covariance of the quantum field theory of tachyons seem to arise from representing the Lorentz group in a too-small space ${\cal F}$, rather than in ${\cal F}\otimes{\cal F}^\star$. 

Delving further into the detailed formulation of our proposal, we consider a real scalar tachyon field characterized by the Klein-Gordon equation with a ``negative square mass'' term:
\begin{equation}
\label{kgequation}
\left(\square- m^2\right)\phi = 0,
\end{equation}
with the standard mode solutions:
\begin{equation}
\label{normalmodes}
u_{\boldsymbol{k}}(t, \boldsymbol{r}) \equiv \frac{1}{(2\pi)^3 2\omega_k} e^{i(\boldsymbol{k}\cdot\boldsymbol{r}-\omega_k t)},
\end{equation}
orthonormal according to the Wronskian $(\psi,\varphi) = (\varphi,\psi)^*  = - (\varphi^*,\psi^*) \equiv i\int_{t=0}\text{d}^3 \boldsymbol{r} \,\psi^*\overleftrightarrow{\partial_t}\varphi$. Note that the dispersion relation $\omega^2 = \boldsymbol{k}^2-m^2$ dictates that the wave vectors $\boldsymbol{k}$ must satisfy the condition $|\boldsymbol{k}|>m$. This requirement mirrors the classical case, in which a tachyon's momentum $\boldsymbol{p}$ must satisfy the condition $|\boldsymbol{p}|>m$ \cite{Dragan2023}. 

Associated with each mode solution $u_{\boldsymbol{k}}(t, \boldsymbol{r})$ is an annihilation operator $\hat{a}_{\boldsymbol k}$ acting in ${\cal F}$ and satisfying the standard commutation relations:
\begin{equation}
\label{eqn:canonical_commutation_relations}
[\hat{a}_{\boldsymbol{k}}, \hat{a}^{\dagger}_{\boldsymbol{l}}] = 2 \omega_k(2\pi)^3\delta^{(3)}(\boldsymbol{k}-\boldsymbol{l}).
\end{equation}
In the orthodox QFT all of these annihilation operators give rise to the field operator:
\begin{align}
\label{singlefockoperator}
    \hat{\phi}(t,\boldsymbol{r}) \equiv \int_{|\boldsymbol{k}|>m}^{} \text{d}^3\boldsymbol{k}\, \left(
u_{\boldsymbol{k}}(t,\boldsymbol{r})\,\hat{a}_{\boldsymbol{k}}  + u^*_{\boldsymbol{k}}(t,\boldsymbol{r})\,\hat{a}^{\dagger}_{\boldsymbol{k}}   \right),
\end{align}
which is a relativistic scalar for normal particles, but fails to be a scalar for the case of tachyons. A Lorentz boost $\Lambda$ that changes the sign of energy $k^0$ of a space-like four-energy $k$, according to $\sgn (\Lambda k)^0 = -\sgn k^0$, transforms the mode solutions in the following way:
\begin{align}
\label{modeboost}
u_{\boldsymbol{k}}(t,\boldsymbol{r})
 \usebox\mypic
u^*_{\boldsymbol{l'}}(t',\boldsymbol{r'}),
\end{align}
with $\boldsymbol{l'}$ being the spatial part of the four-vector $-\Lambda k$. Such a Bogolyubov transformation acts on the annihilation operators according to $\hat{a}_{\boldsymbol{k}} \usebox\mypic\hat{a}^\dagger_{\boldsymbol{l'}}$ in a way that violates their commutation relations \eqref{eqn:canonical_commutation_relations}. For this reason the field operator $\hat{\phi}$ is not a valid relativistic scalar.

In order to restore covariance of the theory and represent the action of a Lorentz boost $\Lambda$ that changes the energy sign, it is essential to consider states belonging to the \superspace\ ${\cal F}\otimes{\cal F}^\star$. Focusing on the scenario portrayed in Fig.~\ref{absorptionemission} we expect such a boost $\Lambda$ to lead to the following transformation of single-particle states of the ${\cal F}\otimes{\cal F}^\star$ \superspace:
\begin{align}
\label{stateboost}
|1_{\boldsymbol{k}}\rangle\otimes\langle 0|  \usebox\mypic|0\rangle\otimes\langle 1_{\boldsymbol{l'}}|.
\end{align} 
This transition corresponds to a similar transformation for creation operators on ${\cal F}\otimes{\cal F}^\star$:
\begin{align}
\label{operatorboost}
\hat{a}^\dagger_{\boldsymbol{k}}\otimes\hat{\mathbb{1}} 
 \usebox\mypic\hat{\mathbb{1}}\otimes\hat{a}^{\star}_{\boldsymbol{l'}},
\end{align} 
where the star represents a dual operator, which acts on the dual space as follows: $\hat{a}^\star_{\boldsymbol{l'}}\langle 0 |  \equiv \langle 1_{\boldsymbol{l'}}|$. Notably, the transformation \eqref{operatorboost} preserves the commutation relations, because:
\begin{equation}
[\hat{a}^{\star\dagger}_{\boldsymbol{k}}, \hat{a}^{\star}_{\boldsymbol{l}}] = 2 \omega_k (2\pi)^3 \delta^{(3)}(\boldsymbol{k}-\boldsymbol{l}).
\end{equation}
Moreover, all Lorentz boosts $\Lambda$ also preserve the twin vacuum state $|0\rangle\otimes\langle 0 | \in {\cal F}\otimes{\cal F}^\star $ of such a theory, defined as: \begin{equation}
\label{twinvacuum}
\hat{a}_{\boldsymbol{k}}\otimes\hat{\mathbb{1}} |0\rangle\otimes\langle 0 | = 0 = \hat{\mathbb{1}}\otimes \hat{a}^{\star\dagger}_{\boldsymbol{k}} |0\rangle\otimes\langle 0 |.    
\end{equation}
This is ensured by the transformation law \eqref{operatorboost}.
The observations outlined above can be formalized through an explicit unitary representation $U\colon \mathrm{SO}^+(1,3) \to \mathrm{U}({\cal F}\otimes{\cal F}^\star)$, which relates the tachyonic twin states in different frames of reference, detailed in the Supplementary Material.

A relativistically covariant scalar field $\hat{\Phi}$ acting in the Hilbert \superspace\ ${\cal F}\otimes{\cal F}^\star$ has the form:
\begin{align}
\label{superfieldoperator}
\boldsymbol{\hat{\Phi}}(t, \boldsymbol{r}) \equiv \frac{1}{2} \left( \hat{\phi}(t,\boldsymbol{r})\otimes\hat{\mathbb{1}} + \hat{\mathbb{1}}\otimes\hat{\phi}^\star(t,\boldsymbol{r})  \right).
\end{align}
Under the Lorentz group's representation $U$, it satisfies: $\boldsymbol{\hat{\Phi}}(\Lambda^{-1} x) = U(\Lambda)^{-1} \boldsymbol{\hat{\Phi}}(x) U(\Lambda) $. In effect, this implies that in the scenario where $\sgn (\Lambda k)^0 = -\sgn k^0$, we have (see the Supplementary Material):
\begin{align}
\label{superoperatorboost}
U(\Lambda)\left(\hat{a}_{\boldsymbol{k}}\otimes\hat{\mathbb{1}} + \hat{\mathbb{1}}\otimes\hat{a}^\star_{\boldsymbol{k}}\right) U(\Lambda)^{-1} =  \left(\hat{a}_{\boldsymbol{l'}}\otimes\hat{\mathbb{1}} + \hat{\mathbb{1}}\otimes\hat{a}^\star_{\boldsymbol{l'}}\right)^\dagger,
\end{align}
with $\boldsymbol{l'}$ being the spatial part of the four-vector $- \Lambda k$. This time the the operators on both sides of \eqref{superoperatorboost} commute with their hermitian conjugates, so the commutation relations are preserved by the transformation. This implies \eqref{operatorboost}, because the alternative transformation ${\hat{a}_{\boldsymbol{k}}\otimes\hat{\mathbb{1}} \usebox\mypic\hat{a}^\dagger_{\boldsymbol{l'}}\otimes \hat{\mathbb{1}}}$ would again violate the commutation relations.

Our extension of the Hilbert space, in which the Lorentz group is represented, defines a theory that does not change the field operator \eqref{superfieldoperator}, preserves the commutation relations, and leaves the vacuum state $|0\rangle\otimes\langle 0| \in \mathcal{F}\otimes\mathcal{F}^\star$ defined by \eqref{twinvacuum} invariant under the group action. This state describes a situation where no tachyons are prepared in the past, nor are there any tachyons detected in the future. In contrast, neither of naive vacuum states, $|0\rangle\in\mathcal{F}$ and $\langle 0|\in\mathcal{F}^\star$, independently admits a covariant description. In this way, we resolve the problems encountered in previous attempts to formulate a relativistic quantum theory of tachyons \cite{Feinberg1967, Arons1968, Dhar1968}.


\noindent {\it ---Quantum field theory in \superspace.} We now direct our focus towards the formalism and properties of the tachyon quantum field theory operating in the Hilbert space ${\cal F}\otimes{\cal F}^\star$. Our primary goal is to calculate the probability amplitudes for scattering processes. 
For this purpose, we need a c-number valued functional on the \superspace. There is a canonical isomorphism of ${\cal F}\otimes{\cal F}^\star$ with the algebra of Hilbert-Schmidt operators, so at least on its sub-algebra of trace-class operators, a good candidate is the trace: $\Tr\colon {\cal F}\otimes{\cal F}^\star \to \mathbb{C}$. For separable {\superstate}s of the form $|\psi\rangle\otimes\langle\xi|\in {\cal F}\otimes{\cal F}^\star $ it yields:
\begin{align}
\label{trace}
\text{Tr}\left(|\psi\rangle\otimes\langle\xi|\right) =\langle\xi|\psi\rangle.	
\end{align}
Given operators $\hat{O}_{1,i}$ and $\hat{O}_{2,i}$ on ${\cal F}$ we construct an operator $\boldsymbol{\hat{O}}$  on ${\cal F}\otimes{\cal F}^\star$ by setting $\boldsymbol{\hat{O}} = \sum_i \hat{O}_{1,i}\otimes \hat{O}^{\dagger \star}_{2,i}$. The corresponding amplitudes are then computed using the identity:
\begin{align}
\label{averagesep}
\text{Tr}\left(\sum_i \hat{O}_{1,i}\otimes \hat{O}^{\dagger \star}_{2,i} |\psi\rangle\otimes\langle\xi|\right) = \langle \xi |	\sum_i \hat{O}_{2,i}^\dagger\hat{O}_{1,i}|\psi\rangle.
\end{align}
This implies that the action of any operator $\boldsymbol{\hat{O}}$ on separable states of the \superspace\ ${\cal F}\otimes{\cal F}^\star$ can be effectively represented using the following operator acting on the Fock subspace ${\cal F}$:
\begin{align} \label{transf_observables}
\sum_i \hat{O}_{1,i}\otimes \hat{O}^{\dagger \star}_{2,i} \to \sum_i \hat{O}^\dagger_{2,i}\hat{O}_{1,i}.
\end{align}

Our approach, deduced from the requirement of Lorentz invariance of QFT of tachyons, bears a remarkable resemblance to the two-state vector formalism in non-relativistic quantum mechanics, developed by Aharonov, Bergmann, and Lebowitz \cite{Aharonov1964}. This formalism describes quantum processes and measurements in a time-symmetric manner, utilizing states that were prepared both in the past and in the future. As shown in \cite{Aharonov1964}, the quantum theory can be reinterpreted using this paradigm, which we will adopt to interpret the states of the \superspace\ ${\cal F}\otimes{\cal F}^\star$.

The two-state vector formalism offers a useful guiding principle for understanding the twin space \cite{Aharonov1984, Aharonov1985, Aharonov2014}. According to that approach, $\mathcal{F}$ contains information on {\it pre-selected} quantum states, i.e., states prepared in the past, while the factor $\mathcal{F}^\star$ corresponds to {\it post-selected} states prepared in the future. In the interaction picture of a free quantum theory in a given reference frame, this correspondance becomes an equivalence, allowing the {\it pre-} and {\it post-selected} states to be identified with the `in' and `out' states, respectively.

The free QFT of tachyons considered in the \superspace\ becomes even more interesting when we apply this formalism to non-separable {\superstate}s, which have no straightforward analogue in conventional QFT. For the most general \superstate\ $\sum_j \alpha_j |\psi_j\rangle\otimes \langle \xi_j| \in {\cal F}\otimes{\cal F}^\star$, we find:
\begin{align}
{\text Tr}	\left( \sum_i \hat{O}_{1,i}\otimes \hat{O}^{\dagger \star}_{2,i} \sum_j \alpha_j |\psi_j\rangle\otimes \langle \xi_j|\right) \nonumber \\
= \sum_j\alpha_j \langle \xi_j |	\sum_i \hat{O}_{2,i}^\dagger\hat{O}_{1,i}|\psi_j\rangle.
\end{align}
This expression cannot be represented in the conventional form given by the right-hand side of \eqref{averagesep} for any states $\langle \xi |$ and $|\psi\rangle$. Notice that if separable states in the space $\mathcal{F}\otimes\mathcal{F}^\star$ become non-separable during evolution, such evolution would be non-unitary within the single space $\mathcal{F}$. In general, it would be represented by a completely positive map.

The possibility of states analogous to our non-separable states within the two-state vector formalism has already been explored by Aharonov and Vaidman \cite{Aharonov1991, Aharonov2007}. Nevertheless, in a Lorentz-invariant QFT the existence of such non-separable states in ${\cal F}\otimes{\cal F}^\star$ is not merely a mathematical curiosity; instead, they must be acknowledged as physically accessible {\superstate}s, as shown in the following example.

Consider a pair of `out' states: $\langle\xi_1|$, $\langle\xi_2|\in{\cal F}^\star$, such that under the representation of a certain Lorentz boost $\Lambda$, the first state remains in the space of `out' states $\langle\xi'_1|\in{\cal F}^\star$, while the second one is transformed onto an `in' state $|\xi'_2\rangle\in{\cal F}$. Then, consider the superposed state: $|0\rangle \otimes \frac{1}{\sqrt{2}}\left(\langle\xi_1| + \langle\xi_2|\right)$, which is separable in ${\cal F}\otimes{\cal F}^\star$. Under the action of the Lorentz boost representation this state becomes non-separable:
\begin{equation}
|0\rangle \otimes \frac{1}{\sqrt{2}}\left(\langle\xi_1| + \langle\xi_2|\right) \xmapsto{U(\Lambda)} \frac{1}{\sqrt{2}}\left( |0\rangle \otimes \langle\xi'_1| +  |\xi'_2\rangle \otimes \langle0|\right).
\end{equation}
This example highlights that non-separable states are physically significant, and their inclusion in the theory is crucial to retain its relativistic invariance. 

Of course, the above construction can be extended to the QFT of subluminal particles, but it is somewhat trivial, as no Lorentz transformation can shift states between $\mathcal{F}$ and $\mathcal{F}^\star$. In this case, we simply end up with two copies of the theory: one for pre-selected states and another for post-selected states.


\noindent {\it ---Interacting theory.} In order to consider an interacting theory, we need to deal more seriously with time evolution and to calculate $S$-matrix elements. To this end, we define two operators:
\begin{align}
\label{hpm}
\hat{\mathbb{H}}_\pm=\hat{H}\otimes\hat{\mathbb{1}}\pm\hat{\mathbb{1}}\otimes\hat{H}^\star,
\end{align} 
with $\hat{H}$ and $\hat{H}^\star$ being the single Fock space Hamiltonians acting in~$\mathcal{F}$ and $\mathcal{F}^\star$, respectively. The operator $\hat{\mathbb{H}}_-$ generates the time evolution of both parts of the twin state in the same direction, i.e., $\mathrm{e}^{-i\hat{\mathbb{H}}_-t}|\psi(0)\rangle\otimes\langle\xi(0)|= |\psi(t)\rangle\otimes\langle\xi(t)|$, so it leaves the amplitudes computed as in \eqref{trace} invariant.

In contrast, the operator $\hat{\mathbb{H}}_+$ generates the time evolution of both parts of the twin state in opposite directions, i.e., $\mathrm{e}^{-i\hat{\mathbb{H}}_+t}|\psi(0)\rangle\otimes\langle\xi(0)|= |\psi(t)\rangle\otimes\langle\xi(-t)|$, thus allowing a construction of an 'in-out' state $|\alpha_\text{in}\rangle\otimes\langle\beta_\text{out}|$ from the eigenstates of the free Hamiltonian $|\alpha_0\rangle\otimes\langle\beta_0|$. Indeed, we have
\begin{equation}
    |\alpha_\text{in}\rangle\otimes\langle\beta_\text{out}|=\lim_{T\to\infty}\mathrm{e}^{-i\hat{\mathbb{H}}_+T}\mathrm{e}^{i\hat{\mathbb{H}}_{0+}T}|\alpha_0\rangle\otimes\langle\beta_0|,
\end{equation}
where $\hat{\mathbb{H}}_{0+}$ denotes the operator corresponding to the free Hamiltonian $\hat{H}_0$. With this, we can write the $S$-matrix element $S_{\alpha\beta}=\langle{\beta_\text{out}|\alpha_\text{in}}\rangle$, in the following way:
\begin{equation}\label{S matrix element}
    S_{\alpha\beta}=\lim_{T\to\infty}\Tr\left(\mathrm{e}^{-i\hat{\mathbb{H}}_+T}\mathrm{e}^{i\hat{\mathbb{H}}_{0+}T}|\alpha_0\rangle\otimes\langle\beta_0|\right).
\end{equation}
The above formula allows us to compute the $S$-matrix elements within the twin space formalism along the same lines as in standard QFT.

It should be emphasized that the operators $\hat{\mathbb{H}}_\pm$ are not derived from the field operator $\boldsymbol{\hat{\Phi}}$. Instead, to obtain them one constructs the single Fock space operator $\hat{H}$ from the operator $\hat{\phi}$ as if it were a Hamiltonian on $\mathcal{F}$, and extends it to the twin space $\mathcal{F}\otimes\mathcal{F}^\star$ like in \eqref{hpm}. This approach results in the covariant expression \eqref{S matrix element} providing a straightforward reference to the standard formulation of QFT.

Using Eq.~\eqref{S matrix element} we can compute the $S$-matrix elements in a perturbative way. It is important to note, however, that the contraction function
\begin{equation}\label{contraction function}
    \langle0|T\hat{\phi}(x)\hat{\phi(y)}|0\rangle=\int_{|\boldsymbol{k}|>m}\frac{\mathrm{d}^4 k}{(2\pi)^4}\frac{i\mathrm{e}^{-ik(x-y)}}{k^2+m^2+i\epsilon}
\end{equation}
is not relativistically invariant because of the restriction $|\boldsymbol{k}|>m$. 
As proposed by Dhar and Sudarshan \cite{Dhar1968} the contraction function can be extended 
by dropping the condition $|\boldsymbol{k}|>m$, which would correspond to including the virtual tachyons with $|\boldsymbol{k}|<m$ into considerations. The propagator obtained in this way is relativistically invariant, as desired. This is similar to the situation in quantum electrodynamics, where (in the Coulomb gauge) we need to include in the propagator the non-physical longitudinal and scalar photons. Just like the longitudinal and scalar photons, the tachyons with $|\boldsymbol{k}|<m$ are allowed to appear as virtual particles, but are excluded from the space of asymptotic states.




As an illustrative example of the $S$-matrix element calculation, let us consider the process depicted in Fig.~\ref{absorptionemission}, where a tachyon is emitted by a subluminal particle governed by the scalar Yukawa Hamiltonian. 
Within the framework of first-order perturbation theory, the $S$-matrix element for this process is given by:
\begin{equation}\label{amplitude1}
    -ig(2\pi)^4\delta^{(4)}(k-l-p),
\end{equation}
where $g$ denotes the coupling constant, and $k$, $l$, and $p$ represent the four-momenta of the subluminal particles in the initial and final states, and the emitted tachyon, respectively.

To study relativistic properties, we analyze this process in a boosted frame (see Fig.~\ref{absorptionemission}). Lorentz transformation $U(\Lambda)$ changes an outgoing tachyon with four-momentum $p$ into an incoming one with $p'=-\Lambda p$. The four-momenta of subluminal particles also shift to $k'=\Lambda k$ and $l' = \Lambda l$. In this boosted frame, we find that the transformed matrix element takes the form:
\begin{equation}\label{amplitude2}
    -ig(2\pi)^4\delta^{(4)}(k'-l'+p') \equiv -ig(2\pi)^4 \delta^{(4)}(\Lambda(k-l-p)),
\end{equation}
which demonstrates the covariance of the scattering process. 

The renormalization procedure for this theory can be carried out at one loop in a standard way, since the UV divergences in diagrams involving tachyons are the same as for scalars with positive mass squared. At the technical level, one needs to subtract singularities from on-shell particles on intermediate lines to avoid double counting and to restrict integration over momenta for tachyons which does not affect logarithmic divergences.

Similar reasoning can be carried out for other types of covariant interactions. Even if some tachyons are boosted from the initial to the final states, this change is compensated by the minus sign of the boosted momentum. As a result, the conditions of momentum conservation at each vertex transform covariantly between all inertial frames.  



\noindent {\it ---Discussion and Conclusions.} We showed how to covariantly quantize a tachyonic field while maintaining the positive-energy spectrum and preserving a stable, Lorentz-invariant vacuum state. Unlike Feinberg \cite{Feinberg1967}, Arons, Sudarshan, and Dhar \cite{Arons1968,Dhar1968}, Schwartz \cite{Schwartz}, as well as others, but similar to Schwartz \cite{Schwartz2018} we proposed to solve this problem by extending the Hilbert space to ${\cal F}\otimes{\cal F}^\star$. We developed an explicitly covariant framework that keeps the commutation relations the same in all reference frames, and it ensures the dynamical stability and relativistic invariance of the vacuum state. We also applied our framework to account for interactions with other fields. 

Our results highlight the existence of a new category of quantum states bearing resemblance to the generalized two-state vectors discussed by Aharonov and Vaidman in \cite{Aharonov1991, Aharonov2007}. But these non-separable states in the \superspace\ ${\cal F}\otimes{\cal F}^\star$ are not merely mathematical curiosities. These states are physically accessible because they can be generated by Lorentz-boosting the conventional, separable states. We argue that our framework singles out the two-vector formalism as a preferred interpretation of the quantum theory, particularly within its relativistic context. However, pinpointing the appropriate non-relativistic quantum mechanical interpretation in light of our results remains an open question, one that necessitates more comprehensive investigation and discourse in future studies.


Numerous other topics require further exploration. The non-separable states that emerge within the structure of the ${\cal F}\otimes{\cal F}^\star$ space bear a striking resemblance to indefinite causal structures \cite{chiribella09} studied in the context of non-classical gravity \cite{Bell4time} and relativistic motion \cite{Dimic, debski2022indefinite}. Furthermore, it is tempting to investigate, if our framework can be formulated in superluminal frames of reference \cite{Dragan2020a} which may turn out more natural to describe the physics of tachyons. 


Finally, an important unanswered question is whether the Higgs field, which can be formally regarded as a tachyon field for small values of the field, can be described within the framework introduced in this study. We believe that our quantization scheme may help to understand and explore the physics of the Higgs phase transition and the dynamics of the broken versus the unbroken phase. One special aspect of these investigations regards CP-violating interactions in the electroweak sector. The measured Higgs mass corresponds to a second-order phase transition, violating Sakharov conditions for dynamical generation of the baryon asymmetry of the Universe (BAU) in electroweak interactions in the Standard Model of particle physics. This prompted many studies about an extended Higgs sector with additional sources of CP violation, which may account for the BAU. Further investigation is needed to explore this idea.


\noindent {\it ---Acknowledgments.} We thank Iwo Białynicki-Birula and Piotr Chankowski for valuable discussions. A.D. thanks Paweł Jakubczyk for irritatingly useful comments and K. T. for encouragement.
K.D. is financially supported by the (Polish) National Science Center Grant 2021/41/N/ST2/01901.

\bibliography{library}

\end{document}